**Quantum Light Nano-Imaging**

*Michael Dapolito[1*], Matthew Fu[1], Fuyang Tay[1,2], Suheng Xu[1], Yuchen Lin[1], Neil Hazra[3], Adam K. Williams[1], Samuel L. Moore[1], Rocco A. Vitalone[1], Jonas Kolker[1], Thomas Cherradi[1], Aaron Holman[1], Thomas P. Darlington[4], Mark E. Ziffer[4], Xavier Roy[2], Sebastian Will[1], Cory R. Dean[1], Mengkun Liu[5], A.J. Millis[1,6], Michael M. Fogler[7], Abhay N. Pasupathy[1], P.J. Schuck[4], D. N. Basov[1*]*

[1]Department of Physics, Columbia University; New York, NY 10027, USA.

[2]Department of Chemistry, Columbia University; New York, NY 10027, USA.

[3]Department of Applied Physics and Mathematics, Columbia University; New York, NY 10027, USA.

[4]Department of Mechanical Engineering, Columbia University; New York, NY 10027, USA.

[5]Department of Physics and Astronomy, Stony Brook University; Stony Brook, NY 11794, USA.

[6]Center for Computational Quantum Physics, Flatiron Institute; New York, NY 10010, USA

[7]Department of Physics, University of California at San Diego; La Jolla, CA 92093, USA

*Corresponding authors: md4108@columbia.edu, db3056@columbia.edu

## Abstract

Entanglement and quantum correlations are central to the physics of quantum materials, yet they have remained notoriously difficult to access experimentally. Accessing these phenomena in solids requires quantum optical probes that operate at the native length and time scales of material excitations, below the diffraction limit of light. Developing the requisite tools has previously been infeasible due to the weak intensities of state-of-the-art quantum light sources and the inefficiency of light coupling in near-field light-matter interactions. In this work, we address these challenges and report the development of a quantum light scattering-type scanning near-field optical microscope (q-SNOM) that enables quantum-optical studies of solid-state systems with nanoscale spatial resolution. As a first demonstration, we visualize the self-interference of single hybrid light-matter polaritons in the prototypical van der Waals semiconductor $MoS_2$. We also introduce a polaritonic time-of-flight metrology that exploits the temporal correlations among entangled photons to observe the quasiparticle propagation dynamics at femtosecond time scales. This work establishes a new experimental paradigm for exploring quantum effects in materials at the nanoscale.

Accessing quantum effects in materials at the nanoscale is crucial for uncovering new physics, developing next-generation quantum sensors, and evaluating material platforms for future quantum technologies. These effects can be investigated by probing material excitations with quantum light at their native nanometer length scales, well below the diffraction limit. Previous diffraction-limited far-field experiments have shown that quantum correlations of entangled photons[1] provide a valuable resource for imaging applications[2–5], for probing quantum correlations and temporal dynamics in materials[6–14], as well as for metrologies below classical noise limits[11,15]. Yet, experimental studies of materials with quantum light remain sparse owing to the challenges posed by low photon fluxes and complex detection schemes. These challenges are further exacerbated in near-field nano-imaging by intrinsically weak light-matter coupling, low collection efficiency, and background light contamination, rendering quantum light nano-imaging experimentally elusive.

Here we report on the development of a quantum light scattering-type scanning near-field optical microscope (q-SNOM) that employs entangled photons to empower novel modalities for nano-imaging and nano-spectroscopy of complex materials. By integrating a scattering-type SNOM (s-SNOM) with entangled photon sources, single-photon detectors, and a custom data acquisition protocol, q-SNOM enables near-field imaging under extremely low light conditions and opens access to previously impossible nanoscale experiments.

As a first application of q-SNOM, we investigate the rich physics of the hybrid light-matter quasiparticles known as polaritons[16,17] hosted by atomically layered van der Waals (vdW) materials. Prominent examples of vdW polaritons include plasmon-polaritons[18–20], phonon-polaritons[21,22], and exciton-polaritons[23–26], which originate from the coupling of light to collective electron oscillations, lattice vibrations, and electron-hole bound states, respectively. Near-field imaging of polaritons enables the visualization of their propagation and interference with tens-of-nanometers resolution, enabling local access to the optical response of quantum materials[27–30]. Extending polaritonic near-field imaging to quantum light thus provides new opportunities to investigate how quantum coherence, and eventually entanglement, manifest in vdW materials.

We demonstrate q-SNOM's capabilities by visualizing the quantum self-interference of single polaritons at their native length and time scales in the prototypical vdW semiconductor $MoS_2$. The quasiparticles we study here are exciton polaritons arising from the hybridization of near-infrared photons with a small admixture of dipole-active excitons. Such polaritonic hybridization gives rise to propagating waveguide (WG) modes[24,27,31] that we launch and visualize, one photon at a time. By directly resolving the self-interference of individual polaritons with tens-of-nanometers spatial resolution, our results establish q-SNOM as a powerful platform for investigating and controlling quantum phenomena in materials at the nanoscale.

## Detecting the single-photon near-field signal

A central challenge in tip-based near-field microscopy is isolating the weak near-field signal from the far-field background. In both classical s-SNOM and q-SNOM, a metallized atomic force microscopy tip (radius ~15 nm) is illuminated by focused light and raster scanned across a sample. The spatial resolution is governed by the tip radius and is unrelated to the wavelength of light. Illumination induces a localized charge distribution at the tip apex that interacts with the sample to generate an enhanced optical near-field. As the tip oscillates at frequency $\omega_{tip}$, the near-field interaction generates tip-scattered light components at integer harmonics of the oscillation frequency ($j\omega_{tip}$). The tip-scattered light ($S_j$) is collected by a parabolic mirror and sent to a detector. In conventional s-SNOM, the detected optical signal is converted into an analog electrical signal, sent to a lock-in amplifier, and demodulated at $j \geq 2$ to suppress the far-field background and isolate the true near-field signal[32–34]. In q-SNOM, however, measurements are performed in the single-photon regime where no analog signal exists. Thus, conventional lock-in detection cannot be employed, necessitating the development of an all-digital demodulation protocol, described below.

The q-SNOM shown in Fig. 1a operates with non-degenerate time-energy entangled photon pairs generated via spontaneous parametric down-conversion (SPDC) in a periodically poled nonlinear crystal. The photon pairs are generated at 1538 nm (806 meV) and 1558 nm (796 meV) with orthogonal polarizations (Supplementary Fig. 1). The emission rate is about $n_{source} \sim 25 \times 10^6\ Hz$. Here, photon pairs are routed into paths $\boldsymbol{P_1}$ and $\boldsymbol{P_2}$. In path $\boldsymbol{P_1}$, the photon impinges on the tip and sample, where it can couple into the sample with probability less than 0.1%. Following interaction with the tip-sample system, the $\boldsymbol{P_1}$ photon is scattered out, collected by a parabolic mirror, coupled to a single-mode fiber, and sent to detector D1, a superconducting nanowire single-photon detector (SNSPD). In path $\boldsymbol{P_2}$, the photon is sent directly through free space to detector D2 to provide timing information between the two photons from the same pair.

Upon reaching the tip, the $\boldsymbol{P_1}$ photon can take three distinct pathways, denoted $\boldsymbol{P_1^A}$, $\boldsymbol{P_1^B}$, and $\boldsymbol{P_1^C}$ in Fig. 1a. In pathway $\boldsymbol{P_1^A}$, the photon is scattered directly by the tip toward the detector without exciting a propagating mode in the sample. In pathway $\boldsymbol{P_1^B}$, the photon launches a polariton that travels in the interior of the crystal toward the sample edge. Upon reaching the edge, the polariton is outcoupled back into a free-space photon and sent to D1 (see Supplementary Section S2 for additional details). Lastly, pathway $\boldsymbol{P_1^C}$ corresponds to a surface mode known as the air radiation mode[23,27]. The coherent superposition of $\boldsymbol{P_1^A}$ and $\boldsymbol{P_1^B}$, whose relative phase varies across the sample surface, gives rise to the single-polariton interference discussed in the next section.

Every time a photon is detected in our setup in Fig. 1a, the SNSPD outputs an electrical pulse that is sent to a time-tagger and timestamped with picosecond resolution. To isolate the near-field interaction, we operate q-SNOM in tapping mode and exploit the tip-induced modulation of the count rate. Specifically, the photon arrival times are binned with the tip-sample distance during each oscillation cycle (see Supplementary Section S3). Figure 1c shows the result of this binning procedure and reveals the single-photon near-field interaction when the tip is at the closest approach to the sample. Recording the tip-modulated single-photon signal, pixel by pixel, allows us to image the near-field interaction with nanometer resolution.

The near-field harmonics of the tip-modulated signal are extracted by Fourier transforming the full tip-cycle histogram (Supplementary Figs. 2b,c). The magnitudes of the Fourier coefficients ($C_j$) yield the amplitudes of the near-field scattering signal[35–37] at each harmonic. We refer to this procedure as digital demodulation.

Detecting both the $\boldsymbol{P_1}$ and $\boldsymbol{P_2}$ photons further enables two additional imaging modalities. Specifically, the $\boldsymbol{P_2}$ photon allows us to track the arrival time delay, $\tau$, between the two photons in paths $\boldsymbol{P_1}$ and $\boldsymbol{P_2}$ as we scan the sample. From the time delays, we extract the second-order correlation function, $g^{(2)}(\tau)$, and the coincidence rate, $n_{coinc}$, at each pixel to generate a single-photon image. A representative $g^{(2)}(\tau)$ trace for our SPDC source is shown in Fig. 1b. Mapping $g^{(2)}(\tau)$ also allows us to observe the ultrafast timing dynamics of propagating polaritons with tens-of-femtoseconds resolution, as discussed in the final section.

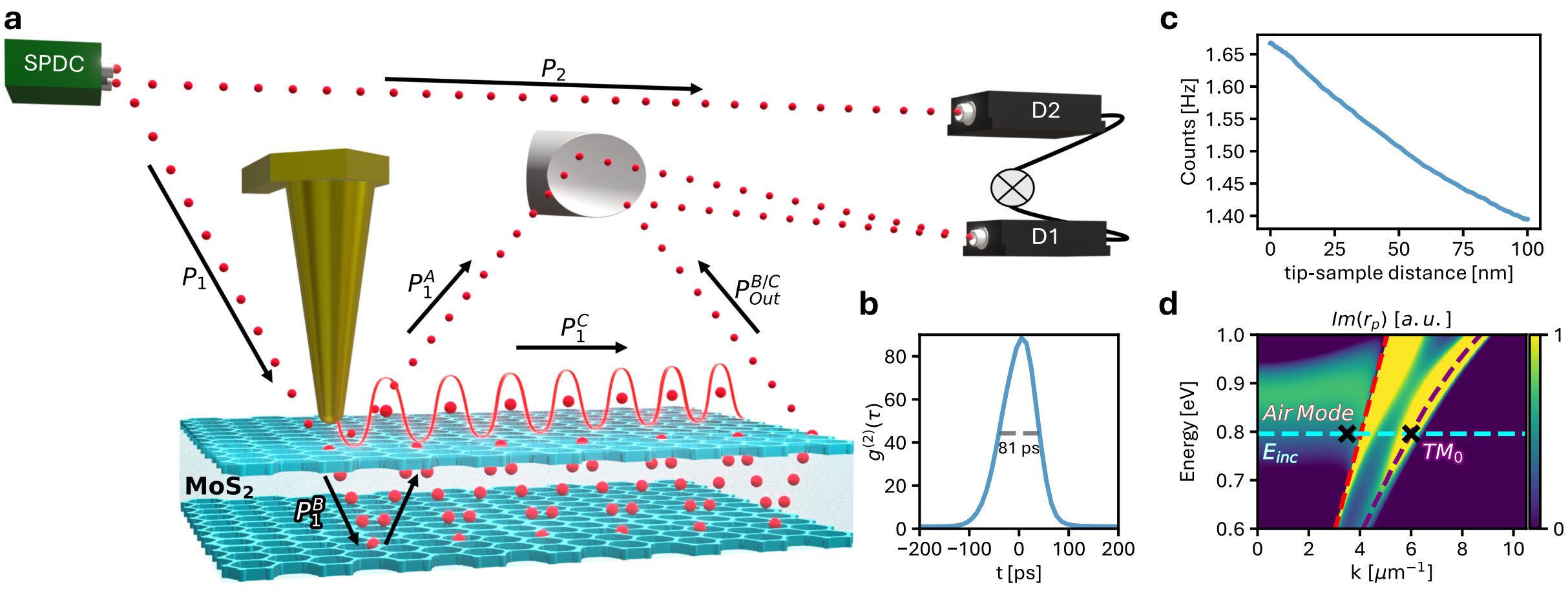

***Fig. 1. Quantum light scattering-type scanning near-field optical microscope (q-SNOM). a,*** *Two photons from an SPDC pair are sent to two paths:* $\boldsymbol{P_1}$ *and* $\boldsymbol{P_2}$*. The* $\boldsymbol{P_1}$ *photon is incident on the tip and can either scatter from the tip (*$\boldsymbol{P_1^A}$*), launch a single polariton in* $MoS_2$ *via the tip-sample interaction (*$\boldsymbol{P_1^B}$*), or scatter from the sample surface (*$\boldsymbol{P_1^C}$*). Exactly one polariton is launched at a time and gets outcoupled back into the far-field at the sample edge. The* $\boldsymbol{P_1}$ *photon interferes with itself via path interference at detector D1. The* $\boldsymbol{P_2}$ *photon is sent directly to detector D2.* ***b,*** *The second-order correlation function,* $g^{(2)}(\tau)$*, of the SPDC source (Supplementary Section S4).* ***c,*** *A single-photon near-field approach curve obtained by correlating the photon arrival times with the tip motion as described in* Supplementary Section S3. ***d,*** *Simulation of the energy-momentum dispersion for the* $MoS_2$ */*$SiO_2$ *heterostructure being excited with p-polarized light. The dispersion is plotted in the form of the imaginary part of the reflection coefficient,* $Im(r_p)$*. Peaks in* $Im(r_p)$ *trace the dispersions of supported modes as a function of excitation energy and mode momentum, k. The red dashed line indicates the free-space light line. The purple dashed line indicates the* $TM_0$ *polaritonic waveguide mode. The horizontal cut at our excitation energy (blue dashed line at 796 meV) reveals the two modes accessible in our experiment: the air radiation mode and the sub-diffractional polaritonic waveguide mode. The* $\times$*-markers indicate the experimentally measured mode momenta.*

## Detecting single polaritons

Polaritonic modes in crystals are uniquely described by their energy-momentum dispersion[38]. Commonly, polaritonic dispersions for transverse magnetic ($TM$) modes are represented by the imaginary part of the reflection coefficient for p-polarized light, $Im(r_p)$[23,39]. Figure 1d shows the calculated $Im(r_p)$ for our $MoS_2/SiO_2$ heterostructure as a function of incident light energy from 0.6-1.0 eV (1.2-2 μm) and mode wavevector $k$. Details of the dispersion calculation are given in Supplementary Section S5. The intersections of the horizontal cut at our probing energy with the poles in $Im(r_p)$ reveal two experimentally accessible modes. The modes correspond to the air radiation mode confined to the sample-air interface and the first-order $TM$ ($TM_0$) WG mode of the $MoS_2$ slab.

Using q-SNOM, we image the polariton mode revealed in $Im(r_p)$ with single-photon sensitivity. By exciting the sample with either a laser or single photons, q-SNOM allows us to image both the classical interference fringes produced by $\sim 10^{15}$ polaritons and the self-interference of single polaritons, respectively.

Figure 2 shows q-SNOM images of polariton fringes (left) and their horizontal spatial derivatives (right) in a 150-nm-thick microcrystal of $MoS_2$ in both the quantum and classical regimes. The quantum light images (Figs. 2a,b) are obtained with one photon from the SPDC pair incident on the tip (rate: $n_{inc} \sim 25$ MHz). The Fig. 2a ("quantum one-arm") image is generated by counting all the photons that arrive at the detector along path $\boldsymbol{P}_1$ ($n_{out,P_1} = 30\ kHz$) and demodulating at the first harmonic ($C_1$). Figure 2b ("coincidence") is generated by measuring the coincidence detection rate between $\boldsymbol{P_1}$ and $\boldsymbol{P_2}$ photons prior to digital demodulation, using a 25-ns coincidence window (Supplementary Section S6). This coincidence metrology rejects non-simultaneous detection events between the two arms and selects only single-photon events occurring in both arms. The quantum light images document our single-photon sensitivity and ability to track the propagation of a single propagating polariton. The non-demodulated photon count rate used to generate Fig. 2a corresponds to just 0.1% of $n_{inc}$. After discarding detected photons without an entangled partner, we find a coincidence rate of $n_{coinc} = 3$ kHz, or just 0.01% of $n_{inc}$.

Figure 2c ("attenuated classical") shows the sample imaged with a 1550-nm (800 meV) laser diode attenuated to the single-photon level. The count rate is set to match that used for the SPDC data, and the signal is digitally demodulated at $C_1$. Unlike the SPDC source, single-polariton excitation cannot be guaranteed or unambiguously identified with the attenuated laser. Figure 2d ("classical") is obtained using the same laser as in Fig. 2c but with 500 $\mu W$ of power incident on the tip. Here the scattered signal is demodulated at the fourth harmonic ($S_4$) with analog demodulation. The classical light s-SNOM images enable

us to classify the modes we visualize with quantum light and uncover the intricacies of quantum light nano-imaging, as discussed below.

Inspecting Fig. 2, we observe differences between the quantum and attenuated classical light images (Figs. 2a-c) and the bright classical light s-SNOM images (Fig. 2d). Notably, the data in Figs. 2a-c were generated by demodulating the scattered signal at the first harmonic of the tip tapping frequency, whereas Fig. 2d was demodulated at the fourth harmonic. In conventional s-SNOM, the scattered signal demodulated at the first or second harmonic typically contains a substantial far-field background that obscures the true near-field signal[40] (see Supplementary Fig. 3). Yet polaritonic near-field fringes are already evident in the first harmonic signal in both the quantum light and attenuated classical light images (Figs. 2a-c). We attribute the ability to resolve genuine near-field patterns in Figs. 2a-c to the far-field background suppression enabled by few-photon illumination and collection of the scattered light through a single-mode fiber (Supplementary Section S7).

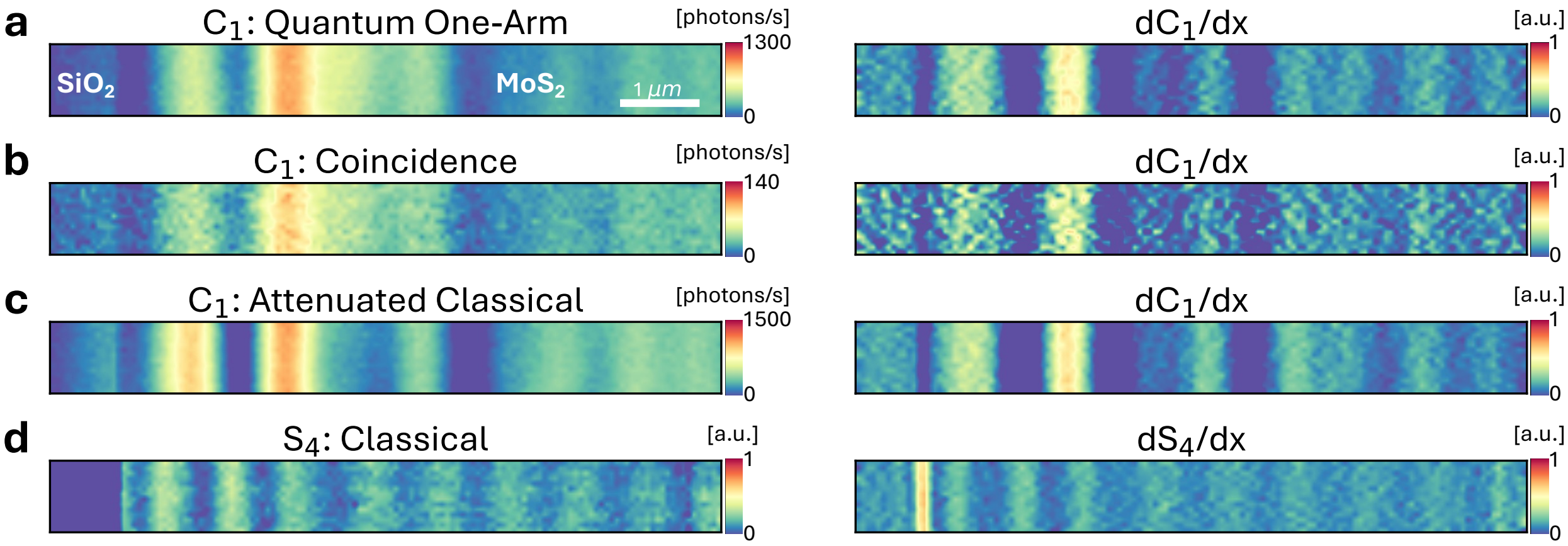

***Fig. 2. Q-SNOM images of polaritonic fringes in $MoS_2$ (left) and their derivatives (right) obtained with quantum and classical light sources. a,*** *Polaritonic WG modes launched via illumination with one photon (1558 nm, 796 meV) from a photon pair at a rate of* $25 \times 10^6$ *photons/s and demodulated at the first harmonic of the tip tapping frequency ($C_1$). Fewer than 1300 photons/s contribute to the first harmonic signal. Both the far-field air mode and the near-field polaritonic mode contribute to the fringe pattern. The integration time is 53.5 s/pixel.* ***b,*** *The same data as in (**a**) after performing a coincidence measurement with a 25-ns window. The coincidence measurement rejects detection events produced by photons without an entangled partner and thus reveals true single polaritons. Fewer than 140 photons/s contribute to the first harmonic signal.* ***c,*** *Polaritonic WG modes excited by a 1550-nm laser diode attenuated to the same count rate as the SPDC source in (**a**). Again,*

*both the air mode and the polaritonic mode contribute to the fringe pattern. The integration time is 40.5 s/pixel.* ***d,*** *Polaritonic WG modes excited by the same 1550-nm laser but with* $500\ \mu W$ *incident on the tip and demodulated at the fourth harmonic ($S_4$). The far-field air mode is absent at this high demodulation order. The integration time is 35 ms/pixel.*

To identify the modes contributing to the data in Fig. 2, we perform a Fourier analysis of the row-averaged derivative images (Fig. 3). The $C_1$ line profiles in Fig. 3a, extracted for the quantum one-arm (green), coincidence (orange), and attenuated classical (blue) cases, reveal a complex beat pattern suggesting contributions from more than one mode. In contrast, the $S_4$ line profile obtained with bright classical light (Fig. 3a,pink) shows a decaying sinusoid expected for interference involving a single mode. The Fourier transforms (FTs) of the line profiles for the quantum and attenuated classical light cases exhibit two distinct peaks centered at angle-corrected momenta $k'_0 = 3.5 \pm 0.1\ \mu m^{-1}$ and $k'_{quant} = 6.0 \pm 0.1\ \mu m^{-1}$. The FT for the bright classical case (Fig. 3b,pink) exhibits a single peak at $k'_{cl} = 6.0 \pm 0.1\ \mu m^{-1}$. The angle-corrected momenta are also represented as ×-markers in Fig. 1d for comparison with the simulated $Im(r_p)$. The markers show that $k'_{quant}$ corresponds to the $TM_0$ WG mode that propagates through the crystal, while $k'_0$ indicates the air radiation mode confined to the sample-air interface. Agreement between the quantum light FTs, bright classical light FTs, and simulated $Im(r_p)$ confirms that we have successfully imaged the near-field polariton mode using single photons.

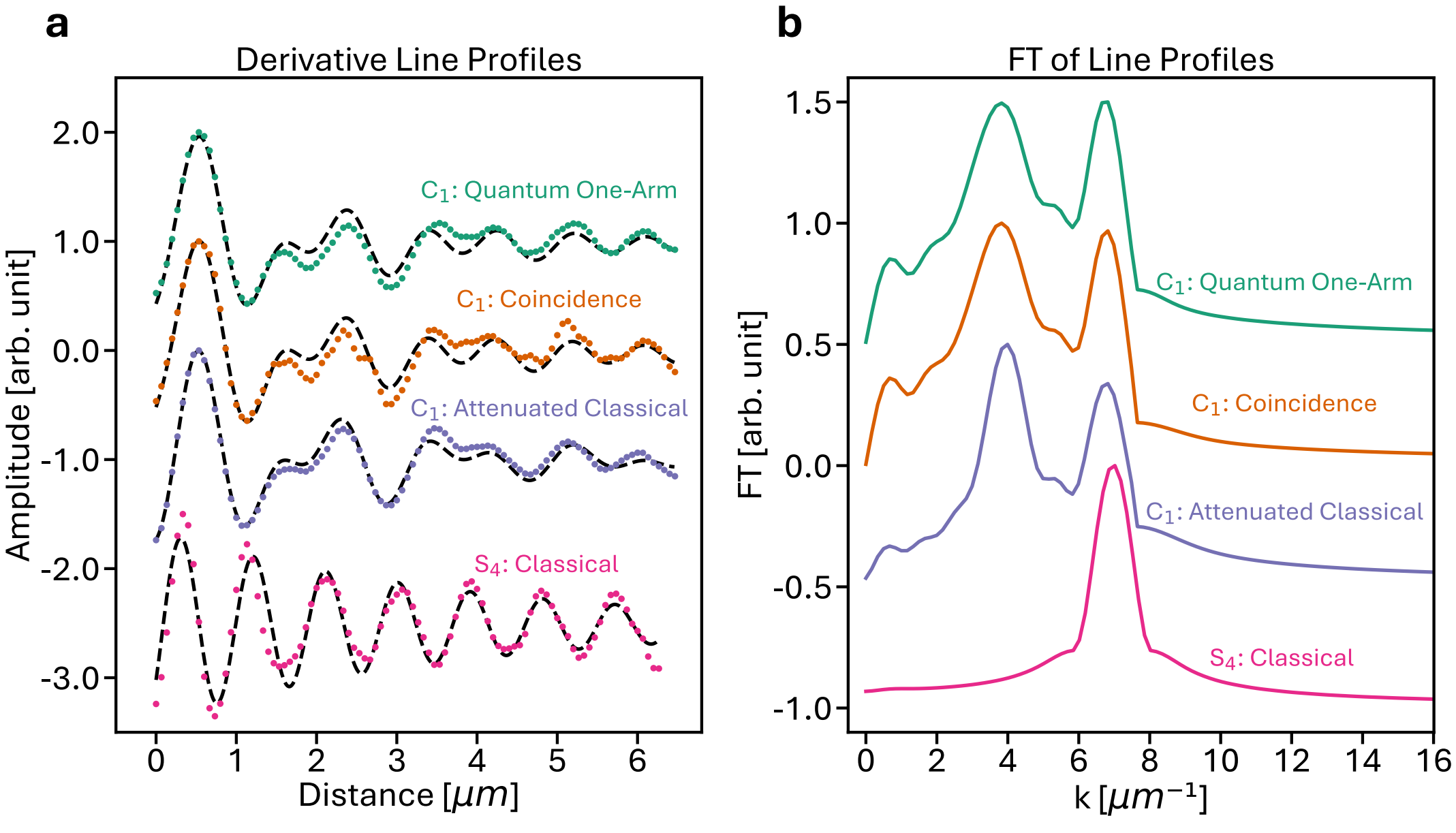

***Fig. 3. Line profiles and Fourier transform analysis of the single- and many-polariton interference patterns. a**, Line profiles obtained by row-averaging the data in Fig. 2,right. Black dashed lines represent the fits described in Supplementary Section S8; colored dots are data points. The line profiles are offset vertically for clarity. **b**, Fourier transforms of the line profile fits in (**a**). The quantum one-arm, coincidence, and attenuated classical FTs reveal peaks corresponding to both the air radiation mode and the $TM_0$ WG mode. The bright classical light FT reveals only the $TM_0$ WG mode. The geometric correction applied to the raw peak positions to obtain the angle-corrected momenta reported in the main text is described in Supplementary Section S9. The matching mode momenta in the quantum and classical cases confirm that q-SNOM visualizes the self-interference of single polaritons with quantum light demodulated at the first harmonic.*

## Single-polariton self-interference

We now discuss the physics of the single-polariton interference. In Fig. 2a, the $\boldsymbol{P_1}$ photons impinge on the tip-sample region once every $t_{inc} = n_{inc}^{-1} = 40\ ns$, whereas the maximum propagation time of the polariton mode through $MoS_2$ is approximately 80 fs. The disparity between these two time scales implies that only one photon is available to participate in any interference pathway at any given time. Furthermore, the coincidence metrology used for Fig. 2b enforces that only one photon is available in path $\boldsymbol{P_1}$ to launch one polariton mode. Figures 2a,b therefore constitute a nanoscale analog of the single-photon double-slit experiment[41,42], realized here in a vdW platform.

In the canonical single-photon double-slit experiment, a photon exists in a coherent superposition of free-space propagation paths. In our nano-optics variant, the photon is prepared in a coherent superposition of three possible pathways: tip-scattering into free space ($\boldsymbol{P_1^A}$), polariton coupling ($\boldsymbol{P_1^B}$), and coupling to the sample surface mode ($\boldsymbol{P_1^C}$). The quantum state may therefore be expressed as $|\psi_{P_1}\rangle = \alpha|P_1^A\rangle + \beta|P_1^B\rangle + \gamma|P_1^C\rangle$, where the complex amplitudes $\alpha$, $\beta$, and $\gamma$ depend on the local tip-sample interaction. Ultimately, the observed fringes originate primarily from interference between the coherent pathways $\boldsymbol{P_1^A}$ and $\boldsymbol{P_1^B}$ and between $\boldsymbol{P_1^A}$ and $\boldsymbol{P_1^C}$. Importantly, pathway $\boldsymbol{P_1^B}$ encodes the phase information accumulated during polariton propagation through the crystal. Therefore, the fringes in Fig. 2 represent a real-space manifestation of single-photon self-interference mediated by nanoscale polaritonic excitations.

## Femtosecond dynamics of single polaritons by time-of-flight nano-imaging

Time-energy entangled photons possess an intrinsic temporal reference due to the simultaneous creation of down-converted photons within a temporal window determined by the pump laser bandwidth and SPDC phase matching[43,44]. We exploit this property to implement time-of-flight imaging of polaritons in $MoS_2$ by mapping $g^{(2)}(\tau)$ at every pixel, thereby tracking the temporal dynamics of each $\boldsymbol{P_1^B}$ photon as it interacts with the sample. The temporal reference is evident in the sample-averaged $g^{(2)}(\tau)$ trace of our SPDC source (Fig. 1b). The $g^{(2)}(\tau)$ trace exhibits a characteristic peak with an $81$-$ps$ width, corresponding to a correlation time of $\sigma_{g^{(2)}} = \frac{81\,ps}{2\sqrt{2\ln|2|}} = 34\,ps$. We increase the temporal resolution beyond the picosecond-scale $g^{(2)}(\tau)$ width, down to $30\,fs$, by averaging the time delay between coincident photons within a narrow coincidence window (see Supplementary Section S10).

The timing dynamics of the polariton signal originate from a time delay, $\Delta\tau$, acquired when a polariton propagates through $MoS_2$ with a slower-than-light group velocity. The acquired time delay, revealed through readings of the temporal separation between $\boldsymbol{P_1^{A/B}}$ photons and their partner $\boldsymbol{P_2}$ photons, is shown schematically in Fig. 4a. When the $\boldsymbol{P_1^A}$ photon scatters off the tip, it propagates through air at the speed of light with a time delay $\tau_0$ relative to the $\boldsymbol{P_2}$ photon. $\tau_0$ is set by the free-space optical path difference between the $\boldsymbol{P_1}$ and $\boldsymbol{P_2}$ arms. However, as the $\boldsymbol{P_1^B}$ photon waveguides through the sample, it travels with group velocity $v_g < c$, resulting in a net time delay $\Delta\tau = \tau_0 + \delta\tau$.

By recording the arrival times of the $\boldsymbol{P_1}$ and $\boldsymbol{P_2}$ photons at every pixel, we acquire $\Delta\tau$ readings for various tip-edge distances, allowing us to map the polariton timing dynamics. Figure 4b shows the averaged line profile of Fig. 2a,right divided into distinct regions (I-VI). Region I contains just the substrate and therefore defines our reference time $\tau_0 = 0\,fs$. The signals in Regions II-VI are formed by both the air mode at $k_0$ and the propagating polariton mode at $k_{TM0}$, resulting in observable $\boldsymbol{P_1}$-$\boldsymbol{P_2}$ time delays. We note that in principle, the sample-surface scattered air mode ($\boldsymbol{P_1^C}$ in Fig. 1a) has a distinct time delay from the tip-scattered light. However, this mode propagates at the speed of light and has a time delay of approximately $\Delta\tau_{air} = 2\pi * (k_0 c)^{-1} = 5\,fs$, which is too small to be observed with our temporal resolution.

Figure 4c shows time-binned traces of the average time delay between coincident photons per pixel for the six regions of Fig. 4b. We use a coincidence window of 25 ps, yielding an average of 5300 coincident photons per pixel. The finite widths of the traces reflect the reduced temporal jitter after time-delay averaging. The time-binned data (circles) in Regions I-VI are fit to double Gaussians (dashed lines) to extract $\Delta\tau$ from the peak positions. The uncertainty in the extracted $\Delta\tau$ values is reduced to $\sigma_{\Delta\tau} = 10\ fs$ because of the additional averaging over the mean coincidence delays (see Supplementary Section S10). Region I produces a trace that peaks at $\tau_0$. The offsets from $\tau_0$ observed in the traces of Regions II-VI result from polaritons traveling progressively longer distances between the tip and the sample edge, and therefore spending more time in the material. The systematic shift of the $\boldsymbol{P_1}$-$\boldsymbol{P_2}$ delay in Regions II-VI constitutes a direct time-domain manifestation of the reduced group velocity of the $MoS_2$ polariton mode. The maximum distance a polariton travels through $MoS_2$ is $7.92\ \mu m$ (Region VI). Figure 4d shows the extracted $\Delta\tau$ values with $3\sigma_{\Delta\tau} = 30\ fs$ error bars, from which the polariton propagation times can be read out. It is observed that the $\boldsymbol{P_1}$ photon spends up to $80\ fs$ traveling through $MoS_2$ when launched from within Region VI.

We verify that the extracted $\Delta\tau$ values are consistent with the polaritonic dispersion and associated group velocity by evaluating the expected polariton propagation time, $\Delta\tilde{\tau}_{v_g}$, for each region. $\Delta\tilde{\tau}_{v_g}$ is calculated from the tip-edge distance and the theoretical polariton group velocity as described in Supplementary Section S10. For comparison, we also calculate the free-space propagation time, $\Delta\tilde{\tau}_c$, for each region. The calculated propagation times are plotted in Fig. 4d, with crosses indicating $\Delta\tilde{\tau}_{v_g}$ and triangles indicating $\Delta\tilde{\tau}_c$. Our measured times agree with the theoretical estimates for all regions within the error bars, affirming the validity of the time-of-flight metrology for tracking polariton propagation in $MoS_2$. The data in Fig. 4 show that q-SNOM provides an ultrafast readout of polariton timing dynamics with a temporal resolution of $30\ fs$. Notably, q-SNOM achieves temporal resolutions typically limited to pump-probe experiments[45–51].

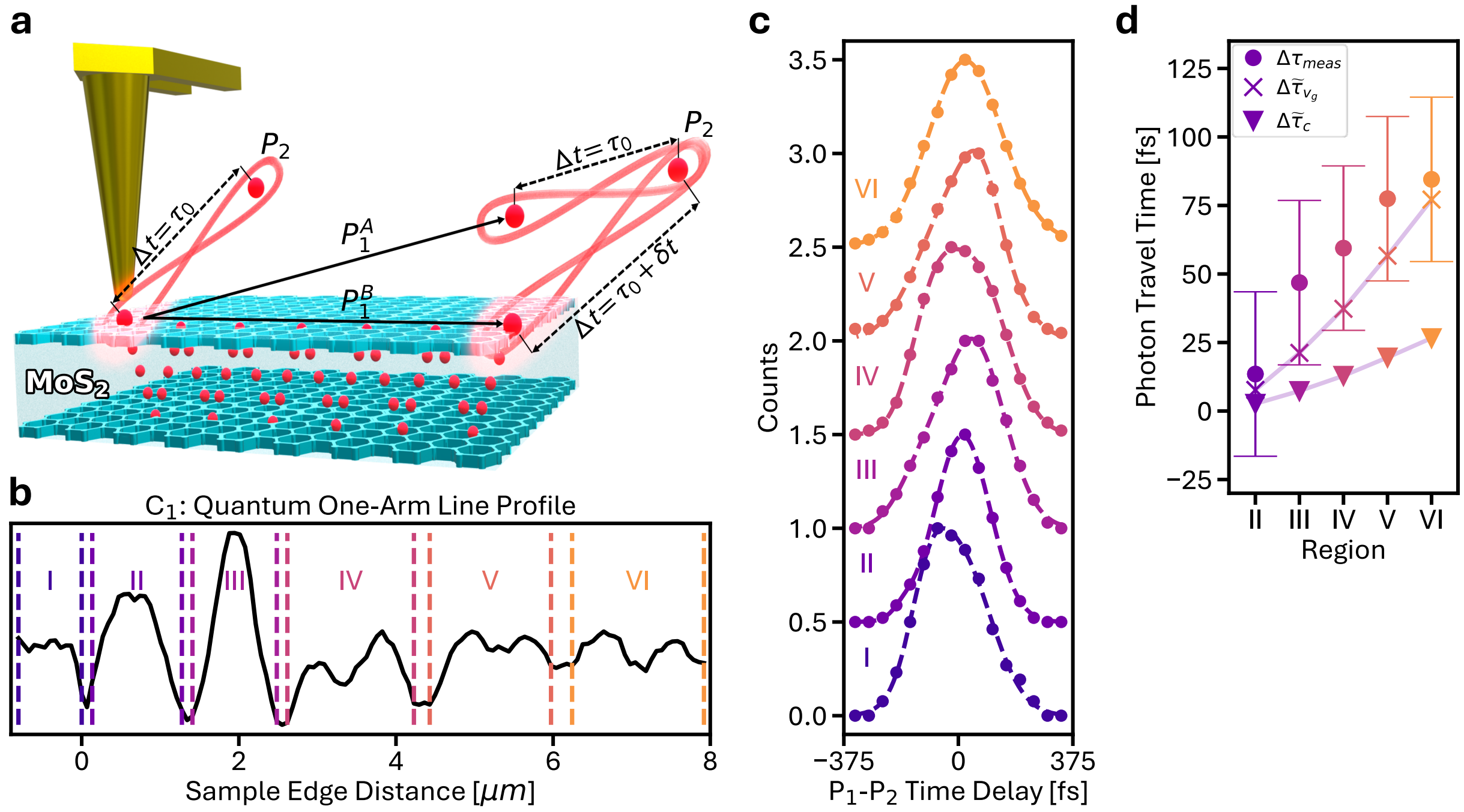

***Fig. 4. Polaritonic time-of-flight experiments empowered by entangled photons. a,*** *Extracting polariton temporal dynamics with q-SNOM nano-imaging. An initial time delay,* $\tau_0$*, arises from the difference in the free-space optical paths of the* $\boldsymbol{P_1}$ *and* $\boldsymbol{P_2}$ *photons. When the* $\boldsymbol{P_1}$ *photon launches a polariton in* $MoS_2$*, it travels with group velocity* $v_g < c$ *and accumulates an additional time delay, Δτ, that depends on the distance traveled.* ***b,*** *The quantum one-arm line profile divided in different regions corresponding to the substrate (Region I) and increasing sample-edge distances (Regions II-VI).* ***c,*** *Binning of the time delay per pixel between coincident* $\boldsymbol{P_1}$ *and* $\boldsymbol{P_2}$ *photons for the different regions in (**b**). Here we use a coincidence window of 25 ps. Dashed lines represent double Gaussian fits described in Supplementary Section S10; circles are the binned data. The polariton timing dynamics are revealed via the small temporal shifts from* $\tau_0 = 0$ *as the photon travels further distances through* $MoS_2$ *to the sample edge.* ***d,*** *Maximum peak positions extracted from the fits in (**c**) compared to theoretical estimates. The polariton propagation times, ranging from 13 to 80 fs, can be read out directly.*

## Outlook

We have developed a quantum light near-field microscope that utilizes entanglement, quantum correlations, and low-light conditions for nano-imaging experiments. The experimental capabilities of q-SNOM were demonstrated by imaging true single-polariton self-interference fringes in space and time in $MoS_2$. Single-polariton self-interference was realized by preparing a photon in a coherent superposition of possible pathways. These results establish q-SNOM as a powerful platform for visualizing single excitations, mapping quantum correlations, and resolving polariton timing dynamics at the nanoscale in quantum materials.

A notable capability of q-SNOM is its ability to access femtosecond-scale dynamics under equilibrium conditions. By using entangled photon partners as an intrinsic timing reference, we directly measured polariton propagation delays without resorting to non-equilibrium pump-probe methods that can substantially modify a system's electronic and lattice properties.

In this study, we employed the most straightforward illumination and detection scheme for imaging with entangled photons. However, q-SNOM can access a wide range of imaging and spectroscopic modalities well suited for probing quantum correlations and coherence in quasiparticles. For example, we envision N00N-state polariton excitation, as well as Hong-Ou-Mandel[52] and Franson[53] interferometric detection schemes. Future work will also explore the self-interference of single plasmons, which contain a large matter component, as well as in situ control of the coherence and amplitude of polaritons via squeezing[54]. Beyond quantum materials, the ultra-low photon flux used for q-SNOM offers opportunities to investigate photosensitive nanoscale biological systems that are inaccessible to conventional near-field microscopy.

## Methods

### Q-SNOM hardware

Q-SNOM utilizes Neaspec's room-temperature neaSCOPE for the standard s-SNOM scanning hardware and software. We use a pixel size of 70 nm and employ a self-homodyne optical detection scheme[34,40]. We use Akiyama probes for our q-SNOM tips.

The SPDC photon source is the correlated narrowband 1550-nm photon pair source from OZ Optics (part # CPS-1000-N-3U3U-1550-9/125), tuned to be slightly non-degenerate. The photon pairs are generated with orthogonal linear polarizations and separated by a polarizing beam splitter inside the source.

Our single-photon detectors are superconducting nanowire single-photon detectors from Single Quantum (Single Quantum Eos 2400 CS). They have 80% quantum detection efficiencies at 1550 nm, $\sim 1\ Hz$ dark count rate, and 13-ps timing jitter.

Our time-tagger is the quTAG standard time-tagger with 3.2-ps RMS jitter and 1-ps resolution.

The tip-based reference pulse required for digital demodulation is generated using an NI PCIe-6361 data acquisition (DAQ) card. The sinusoidal tip motion, read out as a voltage signal from the Akiyama probe, is fed into the DAQ card. The falling edge of the signal triggers the generation of the digital reference pulse.

### Device fabrication

The sample is a 150-nm-thick flake of the transition-metal dichalcogenide 3R-$MoS_2$ exfoliated on a 285-nm-thick $SiO_2$ substrate. The $MoS_2$ was purchased from HQ Graphene.

## Data availability

Example raw timestamp data supporting the findings of this study are available at https://doi.org/10.6084/m9.figshare.32413536. All other data supporting the findings of this study are available from the corresponding authors upon reasonable request.

## Code availability

All code used to process the raw timestamp data supporting the findings of this study is available at https://doi.org/10.6084/m9.figshare.32413536.

## Acknowledgements

The development of the q-SNOM was supported by the National Science Foundation's Major Research Instrumentation (MRI) Program under Grant No. DMR-2408432.

## Author contributions

M.D., M.F., J.K., T.C., A.H., T.P.D., M.E.Z., M.K.L., S.W., P.J.S., A.N.P., and D.N.B. designed and developed the experiment. M.D. performed the measurements. N.H., A.K.W., S.L.M., X.R., and C.R.D. fabricated the samples. M.D., M.F., F.T., S.X., Y.L., N.H., S.L.M., R.A.V., M.M.F., and A.J.M. analyzed the data. M.D., F.T., and N.H. performed the simulations. M.D. and D.N.B. wrote the paper. All authors contributed to the scientific discussions and paper revisions.

## Competing interests

The authors declare no competing interests.